\documentclass[%
 reprint,
superscriptaddress,
 amsmath,amssymb,
prb,
]{revtex4-2}

\usepackage{graphicx}
\usepackage{placeins} 
\usepackage{caption} 
\FloatBarrier 
\usepackage{dcolumn}
\usepackage{bm}
\usepackage{braket}

\usepackage{graphicx}
\usepackage{dcolumn}
\usepackage{bm}

\usepackage{graphicx} 
\usepackage{color}
\usepackage{xcolor}
\usepackage{soul}
\usepackage{float}
\usepackage{booktabs} 
\usepackage{array} 
\usepackage{paralist} 
\usepackage{verbatim} 
\usepackage{subfig} 
\usepackage{float} 
\usepackage{hyperref}
\hypersetup{colorlinks=true,linkcolor=blue,urlcolor=blue}
\begin{document}

\preprint{APS/123-QED}

\title{Floquet non-equilibrium Green's function and Floquet quantum master equation for electronic transport: The role of electron-electron interactions and spin current with circular light}

\author{Vahid Mosallanejad}
\email{vahid@westlake.edu.cn}
\affiliation{Department of Chemistry, Westlake University, Hangzhou, Zhejiang 310024, China $\&$ Institute of Natural Sciences, Westlake Institute for Advanced Study, Hangzhou, Zhejiang 310024, China}

\author{Yu Wang}
\affiliation{Department of Chemistry, Westlake University, Hangzhou, Zhejiang 310024, China $\&$ Institute of Natural Sciences, Westlake Institute for Advanced Study, Hangzhou, Zhejiang 310024, China}

\author{Wenjie Dou}
\email{douwenjie@westlake.edu.cn}
\affiliation{Department of Chemistry, Westlake University, Hangzhou, Zhejiang 310024, China $\&$ Institute of Natural Sciences, Westlake Institute for Advanced Study, Hangzhou, Zhejiang 310024, China}

\date{\today}

\begin{abstract}
Non-equilibrium Green's function (NEGF) and quantum master equation (QME) are two main classes of approaches for electronic transport. We discuss various Floquet variances of these formalisms for transport properties of a quantum dot driven via interaction with an external periodic field. We first derived two versions of the Floquet NEGF. We also explore an ansatz of the Floquet NEGF formalism for the interacting systems. In addition, we derived two versions of Floquet QME in the weak interaction regime. With each method, we elaborate on the evaluation of the expectation values of the number and current operators. We examined these methods for transport through a two-level system that is subject to periodic driving. 
{\color{black}The numerical results of all four methods show good agreement for non-interacting systems in the weak regime}. Furthermore, we have observed that circular light can introduce spin current.
We expect these Floquet quantum transport methods to be useful in studying molecular junctions exposed to light.   
\end{abstract}

\maketitle


\section{Introduction}
Light-matter interactions on quantum scales have long been an attractive research topic\cite{feynman2006qed}. In particular, strong electron-photon interaction opens a new avenue toward engineering material properties \cite{hubener2021engineering,he2022quantum}. In the interpretation of phenomena related to light-matter interaction, the strength of light plays an essential role such that a perturbative treatment of light-matter interaction becomes insufficient \cite{gerry2005introductory}.
Technological advancements, for example, in time-resolved photo spectroscopy \cite{gelin2009efficient}, have allowed us to experimentally demonstrate exotic states such as Floquet-Bloch states on topological insulators \cite{wang2013observation,mahmood2016selective}. Electronic states in such a system can be understood via Floquet theory as if a classical monochromatic light is coupled to a closed fermionic system \cite{faisal1997floquet,fregoso2013driven}. Floquet theory is a non-perturbative mathematical technique that essentially turns the time-dependent schr\"odinger equation with a time-periodic Hamiltonian to a time-independent equation in the price of increasing the dimensionality of the basis set \cite{sambe1973steady,ivanov2021floquet}. 
Floquet theory has been used in the interpretation of Kondo effects, and noise effects \cite{kohler2003controlling,wu2010noise}, spin currents in DNA \cite{rai2013electrically}, electron transfer in donor-bridge-acceptor systems \cite{su2022electron}, 
driven quantum dots \cite{chen2021floquet,gu2023probing}, 
{\color{black}
directional photo-electric effects \cite{kuperman2017field},
}
laser absorption properties\cite{gu2018optical}, nonadiabatic dynamics \cite{wang2023nonadiabatic} and many other applications.

To further enhance our understanding of light-matter interactions, it seems essential to develop approaches that describe how light interacts with open quantum systems. This requirement can be met by properly combining the Floquet theory with quantum transport theories. 
In other words, two classes of quantum transport approaches, namely the non-equilibrium Green's function (NEGF) and the quantum master equation (QME), can be complemented with the Floquet theory.
There have been several works on extending the Floquet theory to open quantum systems. Floquet Green’s function is defined in a number of different ways \cite{faisal1989floquet,stefanucci2008time,tsuji2008correlated,cabra2020optical}. {\color{black}
For example, by employing Floquet Green's function, it is possible to derive an expression (generalized Tien-Gordon) for the current when the baths are subjected to an asymmetric periodic driving \cite{kuperman2015molecular}.}
{\color{black} While the Floquet density matrix method (for closed systems) was established much earlier than the Floquet NEGF \cite{ho1986floquet, grifoni1998driven}, there are some ambiguities in the existence and applicability of Floquet (Lindbladian) QME for open quantum systems \cite{hartmann2017asymptotic,schnell2020there}. 
}
Recent demands for dynamically controlling the material properties by the Floquet engineering stimulated the development of new methods \cite{szczygielski2014application, eissing2016functional, peskin2017formulation, qaleh2022enhancing, engelhardt2019discontinuities,engelhardt2021dynamical,engelhardt2022unified}. 

{\color{black}
The basic NEGF approach is straightforward when a one-body Hamiltonian (e.g., the tight-binding Hamiltonian) can sufficiently describe a system of interest (e.g. in large systems) \cite{datta2005quantum}, and it is not sensitive to the strength of system-environment coupling. More advanced NEGF methods (such as the Hubbard NEGF) are also capable of capturing electron-electron interaction, and other subtle bath-system correlations (e.g. Kondo effects) \cite{cohen2020green}.
Somewhat in contrast, the basic QME approaches (such as the Redfield QME) are often derived under the assumption of weak system-environment coupling and their implementation can be more convenient when inter-dot many-body interactions (e.g. coulomb blockade) are important \cite{elste2005theory}. 
}
Historically, two main flavors for Floquet theory exist in the context of quantum mechanics/solid state physics. The first one that focused on the evolution operator was developed by Shirley in the mid-sixties \cite{shirley1965solution} while the second one that focused on the discreet expansion of wavefunction was developed by Sambe \cite{sambe1973steady}. 
In this work, we intend to discuss that there are multiple ways in which both quantum transport approaches can benefit from the Floquet theory. 

We will offer compact derivations for Floquet non-equilibrium Green's functions (which rely on the expansion of Green's function pioneered by Sambe) and we will also microscopically derive two versions of Floquet quantum master equations (one based on the Floquet-based evolution operator pioneered by Shirley and the other relies on the concept of Floquet density matrix). We will also examine presented methods for a model system.
{\color{black} For the sake of simplicity, we have limited the content of this work to the weak coupling regime. In practice, the validity of QME is limited to the weak coupling regime. As the coupling between the system and bath increases, the outcomes of QME will not match with NEGF formalism, with or without external driving. In addition, we will invoke the wide-band limit (WBL) approximation, which assumes that the density of states of the bath remains approximately constant in the vicinity of the Fermi energy\cite{verzijl2013applicability}. While the wide-band limit (WBL) provides significant simplification for studying quantum transport, for both the QME and NEGF approaches, caution must be exercised as will be clarified in the following.}
To the best of our knowledge, there is no work dedicated to the consistency check between different Floquet quantum transport theories.  
Here, we sketch general Floquet theories which are neither based on an average Hamiltonian approach (Floquet-Magnus) nor based on stroboscopic and micromotion evolution (Van Vleck approximation) \cite{mori2023floquet}. 
{\color{black} While this work does not aim for extreme off-resonance frequencies, we believe the approaches presented here can be applicable across a relatively wide range of driving frequencies.}   
\section{Theory}
\subsection{Model Hamiltonian}
With this work, we are concerned with the electronic transport through a quantum dot (the System) that is weakly connected to the thermal fermionic bath (lead) while it strongly interacts with an external monochromatic light with the frequency $\omega$. In such scenarios, often a time-periodic off-diagonal term (that represents the dipole approximation) should be added to the unperturbed System Hamiltonian ($\hat{H}_S$) such that the total System Hamiltonian will become time-periodic $\hat{H}_S(t)=\hat{H}_S(t+T)$ [$T=2\pi/\omega$]. The bath Hamiltonian, $\hat{H}_B$, and the system-bath interaction, $\hat{H}_{SB}$, will be remained time-independent.
The spinless model Hamiltonian can be given by 
\begin{eqnarray}
\label{eq:1}
\hat{H}_{tot}(t) &=& \hat{H}_S(t)  + \hat{H}_B + \hat{H}_{SB}  
\\
\label{eq:2}
\hat{H}_S(t) &=& \sum_{ij} h_{ij} (t) \hat{d}_i^\dagger \hat{d}_j \\
\label{eq:3}
\hat{H}_B &=& \sum_{l k} \epsilon_{l k} \hat{c}_{l k}^\dagger \hat{c}_{l k} 
\\
\label{eq:4}
\hat{H}_{SB}  &=&  \sum_{l k,i} 
V_{l k, i}  
\hat{c}_{l k}^\dagger \hat{d}_i +\mathrm{H.c.}
\end{eqnarray}
Here, $\hat{d}_i$ ($\hat{d}_i^{\,\dagger}$) is the System's electronic annihilation (creation) operator in the many-body space and $h_{ij}(t)=h_{ij}(t+T)$ represents a periodic one-body Hamiltonian. Likewise, $\hat{c}_{l k}$ ($\hat{c}_{l k}^\dagger$) is the annihilation (creation) operator for the $k$th electronic orbital in the bath $l$.
The quantity $V_{l k, i}$ is the coupling strength between the $k$th orbital of the bath (lead) $l$, and the System's orbital $\hat{d}_i$.
In addition the number operator is $\hat{n}=\sum_{i} \hat{d}_i^\dagger \hat{d}_i$. 
We can further simplify the notation of system-bath interaction Hamiltonian as 
$\hat{H}_{SB} = \sum_{i} 
\hat{C}_{i}^\dagger \hat{d}_i +\mathrm{H.c}$.
We shall also associate the bath $l$ with the electrochemical potential $\mu_{l}$. Note that the System Hamiltonian is quadratic whereas the bath Hamiltonian is non-interacting. 

\subsection{Floquet Green's function}
In the following we explore two types of Floquet Green's function whereby we named 1) Vector-like NEGF, and 2) Matrix-like NEGF. Our starting point for deriving these two flavors of Floquet Green's function is the two-time Kadanoff-Baym (KB) equation \cite{datta2005quantum,jauho1994time} for the retarded Green's function, $G^r(t, t^{\prime})$, as
\begin{eqnarray}
\label{eq:5}
\begin{aligned}
&\Big(i \frac{d}{d t}-h(t)\Big)G^r(t, t^{\prime})-\int_{-\infty}^{+\infty} d t_1 \Sigma^r(t, t_1) \times
\\&
G^r(t_1, t^{\prime})=I \delta(t-t^{\prime}).
\end{aligned}
\end{eqnarray}
Hereafter, we will set $\hbar=1$.
$\Sigma^r=\sum_{l} \Sigma^r_{l}$ refers to the total retarded self-energy. $\Sigma^r_l$ refers to the retarded self-energy of the non-interacting bath $l$, which only depends on the time difference, such that $\Sigma^r(t, t^{\prime})=\Sigma^r(t-t^{\prime})$. 
\subsubsection{Vector-like (V-like) Floquet NEGF}
One way to derive the EOM within the V-like Floquet NEGF is to split the derivation into two parts. In the first part, we define following time-energy retarded and advanced Green's functions
\begin{eqnarray}
\label{eq:6}
\begin{aligned}
&\int_{-\infty}^{\infty} d t^{\prime} G^{r/a}(t, t^{\prime}) e^{i \mathcal{E}(t-t^{\prime})} = G^{r/a}(t, \mathcal{E}),
\end{aligned}
\end{eqnarray}
where $\mathcal{E}$ is the quasi-energy variable. Then, after performing the continuous Fourier transformation, the KB equation for retarded Green's function simplifies as
\begin{eqnarray}
\label{eq:7}
\begin{aligned}
&\Big(\mathcal{E}\,I+i \frac{d}{d t} -h(t)\Big) G^r(t, \mathcal{E})-\int_0^{+\infty} d \tau \Sigma^r(\tau) \times 
\\&
e^{\tfrac{i \mathcal{E}}{\hbar} \tau} G^r(t-\tau, \mathcal{E})= I
\end{aligned}
\end{eqnarray}
Note that $\tau=t-t_1$, and we have employed \textit{convolution} identity in the self-energy term. In the second part, we expand the $G^r(t, \mathcal{E})$ with the following complex Fourier series
\begin{eqnarray}
\label{eq:8}
\begin{aligned}
G^r(t, \mathcal{E})=\sum_{n=-N}^N e^{-i n \omega t} g^r_n(\mathcal{E}).
\end{aligned}
\end{eqnarray}
$N$ is an integer that determines the truncation of the Fourier space basis set. After substitution Eq. (\ref{eq:8}) into Eq. (\ref{eq:7}), we arrive at
\begin{eqnarray}
\label{eq:9}
\begin{aligned}
&\sum_{n}
\left(e^{-i n \omega t} \mathcal{E}~I+e^{-i n \omega t} n
\omega~I-h(t) e^{-i n \omega t}
\right) g^r_n(\mathcal{E})-
\\&
\sum_{n} e^{-i n \omega(t)} \int_0^{+\infty} d \tau \Sigma^r(\tau) e^{i(\mathcal{E}+n \omega) \tau} g^r_n(\mathcal{E})= I.
\end{aligned}
\end{eqnarray}
We then desire to find an EOM for $g^r_n(\mathcal{E})$ which requires multiplying both sides to $e^{i m \omega t}$ and taking the time average over one period. EOM for $g^r_n(\mathcal{E})$ reads as
\begin{eqnarray}
\label{eq:10}
\begin{aligned}
&\Big(
\big(
(\mathcal{E}+m \omega) I
-\bar{\Sigma}^{r}(\mathcal{E}+m \omega)
\big)\delta_{mn}-
\sum_{n} h_{m n}\Big)\times
\\&
g^r_n(\mathcal{E})=\delta_{m 0} I,
\end{aligned}
\end{eqnarray}
where $\bar{\Sigma}^r$ and $[h]_{m n}$ are given by 
\begin{eqnarray}
\label{eq:11}
\bar{\Sigma}^{r}(\mathcal{E}+m \hbar \omega) &=&\int_0^{+\infty} d \tau \Sigma^r(\tau) e^{i(\mathcal{E}+m \omega) \tau}, \\
\label{eq:12}
h_{m n} &=& \frac{1}{T} \int_0^T d t e^{i m \omega t}h(t) e^{-i n \omega t}.
\end{eqnarray}
By running Eq. (\ref{eq:10}) over indices $n$ and $m$, the compact form of V-like Floquet NEGF appears as
\begin{eqnarray}
\label{eq:13}
\begin{aligned}
&\Big(
\mathcal{E}[I^F]
-[h^F]-[\bar{\Sigma}^{r\,F}] \Big)
\mathbf{g}^{r\,F}(\mathcal{E})=\mathbf{I}_o^{F},
\end{aligned}
\end{eqnarray}
where the $[h^F]=\sum_{n,m}h_{m n}-I (m \omega) \delta_{mn}$ is a standard definition for the Floquet Hamiltonian. $[I^F]$ and $[\bar{\Sigma}^{r/a\,F}]$ are block diagonal matrices made of $I$ and $\bar{\Sigma}^{r/a}$.
In the above block matrix EOM, the $\mathbf{g}^{r\,F}$ refers to a stack of $g^r_n$ placed over each other (V-like), and $\mathbf{I}_o^{F}$ refers to a stacking zero matrices symmetrically placed around a central identity matrix. In addition, $\mathcal{E}$ is unbounded.  
{\color{black}Note that with the energy-independent WBL, $\bar{\Sigma}^{r}(\mathcal{E}+m \omega)=\bar{\Sigma}^{r}$, there would be no implementation complexity on the Floquet self-energy term. 
However, for a realistic model caution has to be taken because WBL is valid only for a limited energy window, and the WBL will break as $m\omega$ becomes large (either the frequency or the integer $m$).}
The most important point about Eq. (\ref{eq:13}) is that it is a time-independent EOM.
Such a {\color{black}two-step} procedure was first employed by Sambe to directly solve Schr\"odinger equation with the time-periodic Hamiltonian.  
Also, one can combine the inverse of Eq. (\ref{eq:6}) with Eq. (\ref{eq:8}) to expand the two-time quantity $Q(t, t^{\prime})$ in the V-like form as
\begin{eqnarray}
\label{eq:14}
\begin{aligned}
&Q(t, t^{\prime})=\sum_{n} \int_{-\infty}^{\infty} 
\frac{d \mathcal{E}}{2\pi}
e^{-i(\mathcal{E}+n\omega) t} e^{i \mathcal{E} t^{\prime}} q_n(\mathcal{E}).
\end{aligned}
\end{eqnarray}
From the definition given in Eq. (\ref{eq:14}), the requirement $G^{a}( t^{\prime},t)=G^r(t, t^{\prime})^{\, \dagger}$ indicates $g^a_{n}(\mathcal{E})=g^r_{-n}(\mathcal{E}+n \omega) ^{\,\dagger}$.
\subsubsection{Matrix-like (M-like) Floquet NEGF}
We first define the following one-step transformation for a two-time function
\begin{eqnarray}
\label{eq:15}
\begin{aligned}
&Q(t, t^{\prime})=\sum_{n,m} \int_0^{ \omega} 
\frac{d \mathcal{E}}{2\pi}
e^{-i(\mathcal{E}+n \omega) t} e^{i(\mathcal{E}+m \omega) t^{\prime}} q_{n m}(\mathcal{E}).~~
\end{aligned}
\end{eqnarray}
{\color{black}
Note that, the $Q(t, t^{\prime})$ in Eqs. (\ref{eq:14})-(\ref{eq:15}) is also a function of the driving frequency, however, we have omitted this dependency on the left-hand side to keep the notation simple. 
}
The inverse transformation, for the matrix coefficients $q_{n m}(\mathcal{E})$, is given by
\begin{eqnarray}
\label{eq:16}
\begin{aligned}
q_{n m}(\mathcal{E})=
&\int_{-\infty}^{\infty} 
\frac{d t^{\prime}}{T} \int_0^T d t e^{i(\mathcal{E}+n \omega) t} 
e^{-i(\mathcal{E}+m \omega) t^{\prime}} 
Q(t, t^{\prime}).~~
\end{aligned}
\end{eqnarray}
Contrary to V-like Floquet NEGF, $\mathcal{E}$ is a bounded (independent) variable on the range [$0$ $\omega$]. We proceed by expanding $G^{r}$ using Eq. (\ref{eq:15}) and substituting the result into the KB equation, and solving for the matrix coefficient $g^r_{nm}$. In this process, one can first notice that RHS can be reduced to $I \delta_{n m}$. After some lengthy algebra, we arrive at the following EOM for the retarded Green’s
\begin{eqnarray}
\label{eq:17}
\begin{aligned}
&\Big(
(\mathcal{E}+n \omega) -
\bar{\Sigma}^{r}
(
\mathcal{E}+n\omega
) 
\Big)
g^r_{n m}(\mathcal{E})-
\\&
\sum_{k}h_{n k}~g^r_{k m}(\mathcal{E})=I \delta_{n m}.
\end{aligned}
\end{eqnarray}
The definition of $\bar{\Sigma}^{r}$ and $h_{n k}$ is identical to what is given in Eqs. (\ref{eq:11}) and (\ref{eq:12}). Comparing Eq. (\ref{eq:10}) with Eq. (\ref{eq:17}), it is evident why this form of EOM can be referred to as the Matrix-like Floquet NEGF. The compact form of M-like Floquet Green's function reads as
\begin{eqnarray}
\label{eq:18}
\begin{aligned}
&\Big(
\mathcal{E}[I^F]
-[h^F]-[\bar{\Sigma}^{r\,F}] \Big)
[G^{r\,F}](\mathcal{E})=[I^{F}].
\end{aligned}
\end{eqnarray}
{\color{black}
In the M-like framework, the condition $G^{a}( t^{\prime},t)=G^r(t, t^{\prime})^\dagger$ imposes 
$g^a_{n m}(\mathcal{E})=g^r_{m n}(\mathcal{E})^\dagger$ with no shift in the quasi-energy
which in turn leads to $[G^{a\,F}](\mathcal{E})=[G^{r\,F}](\mathcal{E})^{\,\dagger}$. 
Note that in contrast for a V-like coefficient, $g^a_{n}(\mathcal{E})\neq g^r_{n}(\mathcal{E})^\dagger$, and hence $\mathbf{g}^{a\,F}(\mathcal{E})^{\,\dagger}\neq\mathbf{g}^{r\,F}(\mathcal{E})^{\,\dagger}$.
}
\subsection{Observable in Floquet NEGF}
To explore the transport properties, the expectation values of two (operators) observable are essential. 1) the number operator $\langle \hat{n} \rangle (t)$, and 2) the terminal current 
{\color{black} $\langle \hat{J} \rangle (t)$.}
For $\langle \hat{n} \rangle (t)$, we shall trace the two-time lesser Green's function as $\langle \hat{n} \rangle(t)=-i \operatorname{Tr}(G^{<}(t,t))$ \cite{jauho1994time}. The $G^{<}(t,t^{\prime})$ given by
\begin{eqnarray}
\label{eq:19}
\begin{aligned}
G^{<}(t,t^{\prime})=
\int_{-\infty}^{+\infty} d t_1 \int_{-\infty}^{+\infty} d t_2 
&G^r(t, t_1)
 \times\\
&
\Sigma^{<}(t_1-t_2) G^a(t_2, t^{\prime}).
\end{aligned}
\end{eqnarray}
The total lesser self-energy is denoted by: $\Sigma^{<} = \sum_{l} \Sigma^{<}_{l}$. 
{ 
\color{black}
For $\langle \hat{J} \rangle (t)$, we shall trace the two-time particle current as $\langle \hat{J} \rangle(t)= \operatorname{Tr}({J}(t, t))$.
We define the two-time particle current per spin passing through the contact $l$ as}
\begin{eqnarray}
\label{eq:20}
\begin{aligned}
{J}_l(t, t^{\prime})=
\int_{-\infty}^{+\infty} d t_1
{\color{black} \mathrm{2Re}}
\Big(
&
G^r(t,t_1)
\Sigma_{l}^{<} (t_1,t^{\prime})+
\\&
G^{<}(t, t_1)
\Sigma_{l}^a (t_1,t^{\prime})
\Big).
\end{aligned}
\end{eqnarray}
{\color{black} Note that the above relation is equivalent to the Jauho-Wingreen-Meir expression and valid for interacting devices \cite{stefanucci2004time, wu2010noise}.}
In the following (without details of derivation), we summarize general expressions for Floquet components of these two observable within the V-like and M-like Floquet NEGF frameworks. 

\textit{Observable in the V-like Floquet NEGF.--} 
The expansion given in Eq. (\ref{eq:14}) will be used to expand $G^{</r/a}$ (in terms of $g_n^{</r/a}$) and resulting expressions will be substituted in Eq. (\ref{eq:19}). Then, we proceed to derive the matrix coefficients $g_n^{<}(\mathcal{E})$. After doing some algebra, it reads as
\begin{eqnarray}
\label{eq:21}
\begin{aligned}
&g_n^{<}(\mathcal{E})=\sum_{m}
g_{n-m}^r(\mathcal{E}+m\omega)
\Sigma^{<}(\mathcal{E}+m\omega)
g_m^a(\mathcal{E})=
\\
&
\sum_{m}
g_{n-m}^r(\mathcal{E}+m\omega)
\Sigma^{<}(\mathcal{E}+m\omega)
g_{-m}^r(\mathcal{E}+m\omega)^\dagger.
\end{aligned}
\end{eqnarray}
In the energy domain, the lesser self-energy is: $\bar{\Sigma}^{<}_{l} (\mathcal{E})=i
\Gamma^r_{l}(\mathcal{E}) f_l(\mathcal{E})$, $\Gamma^r_{l}=2 \operatorname{Im}(\bar{\Sigma}^r_{l})$. Here, $ f_l(\mathcal{E})\equiv f(\mathcal{E},\mu_{l})$ and $\mu_l$ are the Fermi function and the electrochemical potential associated with the terminal $l$. {\color{black} We remind that the Fermi function depends on the temperature as well.} 

For the \textit{current}, we will follow the same procedure with expanding the ${J}_l(t, t^{\prime})$ in terms of the matrix coefficient $J_{l,n}(\mathcal{E})$. Solving for $J_{l,n}(\mathcal{E})$, we arrive at 
\begin{eqnarray}
\label{eq:22}
\begin{aligned}
J_{l,n}(\mathcal{E})=
{\color{black} \mathrm{2Re}}
\Big(
&
g_n^r(\mathcal{E})
\bar{\Sigma}_{l}^{<} (\mathcal{E})+
g_n^{<}(\mathcal{E})\bar{\Sigma}_{l}^a (\mathcal{E})
\Big).
\end{aligned}
\end{eqnarray}
\textit{Observable in the M-like Floquet NEGF.--} Similar to above, one can employ Eq. (\ref{eq:15}) to expand $G^{</a/r}$ and substitute the resulting expressions in Eq. (\ref{eq:19}). After doing some algebra, we arrive at the following relation for $g^<_{nm}(\mathcal{E})$
\begin{eqnarray}
\label{eq:23}
\begin{aligned}
&g^<_{nm}(\mathcal{E})=
\sum_{k}
g_{nk}^r(\mathcal{E})
\bar{\Sigma}^{<}(\mathcal{E}+m\omega)
g_{km}^a(\mathcal{E}).
\end{aligned}
\end{eqnarray}
For the \textit{current}, ${J}_{l}(t,t^{\prime})$ has to be expanded in terms of the coefficient matrix $J_{l,nm}(\mathcal{E})$ using Eq. (\ref{eq:15}). Substituting expanded forms of all components into Eq. (\ref{eq:19}) and solving for $J_{l,nm}(\mathcal{E})$ results in
\begin{eqnarray}
\label{eq:24}
\begin{aligned}
J_{l,nm}(\mathcal{E})=
{\color{black} \mathrm{2Re}}
\Big(
g_{nm}^r(\mathcal{E})
\bar{\Sigma}_{l}^{<} (\mathcal{E}+m\omega)+
&\\
g_{nm}^{<}(\mathcal{E})
\bar{\Sigma}_{l}^a (\mathcal{E}+m\omega)
\Big).
\end{aligned}
\end{eqnarray}
Eqs. (\ref{eq:21}-\ref{eq:24}) may look complex at first glance, however, these equations can be further simplified when we evaluate the time averaged of an observable. 
\subsection{Time-average of an observable} 
Within the V-like Floquet NEGF framework, the time average of an observable over one period can be obtained by changing the $t^{\prime}$ to $t$ in Eq. (\ref{eq:14}) and performing the time average over one period. Using Eq. (\ref{eq:21}), the time average of the number operator over one period, $\overline{ n }$, in the V-like framework read as 
\begin{eqnarray}
\label{eq:25}
\begin{aligned}
\overline{ n }
&=
\frac{-i}{T} \int_0^T d t~ \operatorname{Tr}(G^{<}(t, t))=
-i 
\int_{-\infty}^{\infty} 
\frac{d \mathcal{E}}{2\pi}
\operatorname{Tr}(g^<_0(\mathcal{E}))
\\&
=
-i 
\int_{-\infty}^{\infty} 
\frac{d \mathcal{E}}{2\pi}
\sum_{m}
\operatorname{Tr}
\Big(
g_{m}^r(\mathcal{E})
\bar{\Sigma}^{<} (\mathcal{E}+m\omega)
g_m^r(\mathcal{E})^\dagger
\Big).
\end{aligned}
\end{eqnarray}
In the above expression, we have used the property 
$g_{-m}^r(\mathcal{E}+m\omega)
=g_m^r(\mathcal{E})$, and we have also employed the wide-band approximation as well.
We stress that only the central component ($n=0$) of $\mathbf{g}^{<}$ is required to evaluate the time average of $\langle \hat{n} \rangle(t)$ over one period. 

Within the M-like Floquet NEGF framework, the time average of the number operator  over one period reads as
\begin{eqnarray}
\label{eq:26}
\begin{aligned}
\overline{ n}
&
=
-i
\sum_{n} \int_0^{\omega} 
\frac{d \mathcal{E}}{2\pi} 
\operatorname{Tr}(g^<_{n n}(\mathcal{E}))
\\&
=
-i
\int_0^{ \omega} 
\frac{d \mathcal{E}}{2\pi}
\operatorname{Tr}
\Big(
[G^{rF}](\mathcal{E})
[\bar{\Sigma}^{<F}](\mathcal{E})
[G^{rF}]^\dagger(\mathcal{E})
\Big).
\end{aligned}
\end{eqnarray}
Here, $[\bar{\Sigma}^{<F}](\mathcal{E})$ denotes a block diagonal matrix made of $\bar{\Sigma}^{<}(\mathcal{E}+m\omega)$.
Next, we simplify current expressions. In the V-like framework, the time average of the current over a period, $\overline{{J}}$, simplifies as
\begin{eqnarray}
\label{eq:27}
{\color{black}
\begin{aligned}
\overline{{J}}_{l}
&=
\frac{1}{T} \int_0^T d t~ \operatorname{Tr}(J_l(t, t))
=
\int_{-\infty}^{\infty} 
\frac{d \mathcal{E}}{2\pi}
\operatorname{Tr}(J_{l,0}(\mathcal{E}))=
\\&
\int_{-\infty}^{\infty} 
\frac{d \mathcal{E}}{2\pi} 
\operatorname{Tr}
\Big(
\mathrm{2Re}
\big( g_{0}^r(\mathcal{E})
\Sigma_{l}^{<} (\mathcal{E})+g_{0}^{<}(\mathcal{E})
\Sigma_{l}^a (\mathcal{E})
\big)
\Big).
\end{aligned}
}  
\end{eqnarray}
In the M-like framework, based on Eqs. (\ref{eq:15}) and (\ref{eq:24}), the time average of the current over one period read as
\begin{eqnarray}
\label{eq:28}
\begin{aligned}
&
\overline{{J}}_{l}
=
\sum_{n} \int_0^{\omega} 
\frac{d \mathcal{E}}{2\pi} 
\operatorname{Tr}
(J_{l,nn}(\mathcal{E}))=
\\&
\int_0^{ \omega} 
\frac{d \mathcal{E}}{2\pi}
\operatorname{Tr}
{\color{black}
\Big(
 \mathrm{2Re}
 }
\big(
[G^{rF}](\mathcal{E})
[\bar{\Sigma}_{l}^{<F}](\mathcal{E})
+
[G^{<F}](\mathcal{E})
[\bar{\Sigma}_{l}^{aF}]
\big)
\Big).
\end{aligned}
\end{eqnarray}
In the absence of electron-electron interaction, same as in non-Floquet scenarios, the above relation can be further simplified to a Landauer-like expression for a two-terminal device as
\begin{eqnarray}
\label{eq:29}
\begin{aligned}
& 
\overline{{J}}_{l}
=
\int_0^{ \omega} 
\frac{d \mathcal{E}}{2\pi}
T_{ll^\prime}(\mathcal{E})
\big(
 f_l-f_{l^\prime}
\big),
\\&
T_{ll^\prime}(\mathcal{E})=\operatorname{Tr}
\Big(
[\bar{\Gamma}_{l}^{rF}]
[G^{rF}](\mathcal{E})
[\bar{\Gamma}_{l^\prime}^{rF}]
[G^{rF}]^\dagger (\mathcal{E})
\Big),
\end{aligned}
\end{eqnarray}
where $T_{ll^\prime}(\mathcal{E})$ is the two terminal transmission function.  Here, $[\bar{\Gamma}_{l/l^\prime}^{rF}]$ is the block diagonal matrix made of $\bar{\Gamma}_{l/l^\prime}^{r}$.
\subsection{Floquet Green's function for interacting systems}
For interaction systems, we introduce an ansatz for the M-like Floquet Green's function related to the \textit{equation-of-motion} method \cite{meir1991transport,haug2007quantum}.
We should consider the spin degree of freedom in the total Hamiltonian, Eqs. (\ref{eq:1})-(\ref{eq:4}), by modifying annihilation (creation) operators $\hat{d}_{i}^{(\dagger)}/\hat{c}_{i}^{(\dagger)}
\rightarrow \hat{d}_{i,\sigma}^{(\dagger)}/\hat{c}_{i,\sigma}^{(\dagger)}$ in which $\sigma=\uparrow,\downarrow$. The (time-independent) electron-electron interaction, $\hat{H}_{ee}$, is given by a Hubbard-like two-body interaction as
\begin{eqnarray}
\label{eq:30}
\begin{aligned}
\hat{H}_{ee} &=
\sum_{i}
u_i \,
\hat{d}_{i,\sigma}^\dagger \hat{d}_{i,\sigma} 
\hat{d}_{i,\tilde{\sigma}}^\dagger \hat{d}_{i,\tilde{\sigma}}. \\
\end{aligned}
\end{eqnarray}
Here, $\tilde{\sigma}$ refers to opposite spin of $\sigma$. In case of time-independent $h_{ij}$ (Non-Floquet case) and with truncating to double-particle Green's function, 
{\color{black}
$G^{(2)r}(E)$ ($E$ is the energy associated with the Fourier transform),
}
one can obtain a close form for the retarded Green's function as
{\color{black}
\begin{eqnarray}
\label{eq:31}
&
\big(
{E}[I]-[h]-[{\Sigma}^{r}_{\sigma}] 
\big)
[G^{r}_{\sigma}]({E})=[I^{}]+[U][G^{(2)r}_{\sigma}]({E}),~~~
\\&
\label{eq:32}
\big(
{E}[I]-[h]-[U]-[{\Sigma}^{r}_{\sigma}] 
\big)
[G^{(2)r}_{\sigma}]({E})=[\langle n_{\tilde{\sigma}}\rangle],~~~~~
\end{eqnarray}
}
where, $[U]$ refers to a diagonal matrix consist of the elements $\{u_i\}$ \cite{ryndyk2016theory}. The matrix $[\langle n_{\tilde{\sigma}}\rangle]$ is a diagonal matrix consist of occupation numbers $\langle n_{i\tilde{\sigma}}\rangle$. 
{\color{black}
Occupation numbers are related to the lesser Green's functions by
\begin{eqnarray}
\label{eq:23_R2}
\langle n_{{\sigma}/\tilde{\sigma}}\rangle
=-i 
\int_{-\infty}^{\infty} 
\frac{dE}{2\pi}
\operatorname{Tr}\big(G^<_{{\sigma}/\tilde{\sigma}}(E)\big).
\end{eqnarray}
Also, truncating to double-particle lesser Green's function, $G^{(2)<}(E)$, the lesser Green's functions can be obtained by 
\textcolor{black}{
\begin{eqnarray}
\label{eq:34_R2}
&
\begin{aligned}
&\big(
{E}[I]-[h]-[{\Sigma}^{r}_{\sigma}]
\big)
[G^{<}_{\sigma}]=
[U][G^{(2)<}_{\sigma}]+
[\Sigma^{<}_{\sigma}][G^{a}_{\sigma}],
\end{aligned}
\\
\label{eq:35_R2}
&
\begin{aligned}
&\big(
{E}[I]-[h]-[U]-[{\Sigma}^{r}_{\sigma}] 
\big)
[G^{(2)<}_{\sigma}]= 
[\Sigma^{<}_{\sigma}][G^{(2)a}_{\sigma}],
\end{aligned}
\end{eqnarray}
where $\Sigma^{<}_{\sigma}=i \sum_{l}
\Gamma^r_{l}({E}) f_l({E})$.
} 
{\color{black}
Eqs. (\ref{eq:31})-(\ref{eq:35_R2}) provide the simplest improvement to the mean-field approximation for the multi-level Anderson impurity system. Indeed, Eqs. (\ref{eq:31})-(\ref{eq:35_R2}), draw a self-consistent procedure. 
However, (numerically) it can be shown in weak coupling and low-temperature limits that the assumption $\langle n_{i{\sigma}/\tilde{\sigma}}\rangle=1/2$ in Eq. (\ref{eq:32}) [which makes the problem non-self-consistent]  
}
gives a good quantitative prediction on the onsets of current plateaus (charging effects) related to the coulomb blockade phenomenon. 
{\color{black}
However, Eqs. (\ref{eq:31})-(\ref{eq:35_R2}) can not correctly predict correct heights of the quantized plateaus or Kondo peak
and a better scheme is required \cite{galperin2007inelastic}.
}

In previous sections where we discussed the Floquet transformation, we saw that the time-independent components of Green's function equation turn into fixed diagonal blocks. Here, we extend this observation to the interacting case, and hence, we expect the following M-like Floquet version for Eqs. (\ref{eq:31})-(\ref{eq:32}) as
\begin{eqnarray}
\label{eq:33}
&\big(
\mathcal{E}[I^F]
-[h^F]-[\bar{\Sigma}^{rF}_{\sigma}] \big)
[G^{rF}_{\sigma}]=[I^{F}]+[U^F][G^{(2)rF}_{\sigma}],~~~~~~~
\\
\label{eq:34}
&\big(
\mathcal{E}[I^F]
-[h^F]-[U^F]-[\bar{\Sigma}^{rF}_{\sigma}] \big)
[G^{(2)rF}_{\sigma}]=[\langle n_{\tilde{\sigma}}\rangle ^F].~~~~~~~~~~
\end{eqnarray}
{\color{black} 
In addition, and in the same way, we would have the following expressions for the M-like Floquet lesser Green's functions
}
\textcolor{black}{
\begin{eqnarray}
\label{eq:35}
\begin{aligned}
&\big(
\mathcal{E}[I^F]
-[h^F]-
[\bar{\Sigma}^{<F}_{\sigma}]
\big)
[G^{<F}_{\sigma}]=
[U^F][G^{(2)<F}_{\sigma}]+~~~~
\\& [\Sigma^{<F}_{\sigma}][G^{aF}_{\sigma}],
\end{aligned}
\\
\label{eq:36}
\begin{aligned}
&\big(
\mathcal{E}[I^F]
-[h^F]-[U^F]-[\bar{\Sigma}^{rF}_{\sigma}] 
\big)
[G^{(2)<F}_{\sigma}]= ~~~~~~~~~~~~
\\&
[\Sigma^{<F}_{\sigma}][G^{(2)aF}_{\sigma}].
\end{aligned}
\end{eqnarray}
}
We stress that the lesser Green's function for the interacting case is different than the non-interacting one. Consequently, the expectation value of the number operator and the average of the number operator over one period should follow Eq. (\ref{eq:35}). However, the expression for the average current per cycle is identical to the non-interacting case. It is worth mentioning that the Landauer formula is not correct in the case of interacting systems and hence there would be no Landauer-like expression for the average current per cycle.

\subsection{Floquet Quantum Master Equation }
In the following we explore two types of Floquet Quantum Master Equation (FQME) whereby we named 1) Hilbert-Space FQME (HS-FQME), and 2) Floquet-Space FQME (FS-FQME). 
Our starting point for deriving HS-FQME is the well-known Redfield QME in the Schr\"odinger picture whereas we will start from the Floquet Liouville von-Neumann (FLvN) equation for deriving HS-FQME.  
\subsubsection{Hilbert-Space FQME}
When the Hamiltonian of the System is time-dependent (not restricted to the time-periodic), the dynamics of the System's density operator can be notified by the following Redfield equation
\begin{eqnarray}
\label{eq:37}
&
\begin{aligned}
\frac{\partial \hat{\rho}_S(t)}{\partial t}
&=-i\left[\hat{H}_S(t),
\hat{\rho}_S(t)\right]-
\\&
\int_0^{\infty} \operatorname{Tr}_B
\left[
\hat{H}_{SB},
\left[\tilde{\tilde{H}}_{SB}(t, \tau), \hat{\rho}_S(t) \otimes \hat{\rho}_B
\right]\right] d \tau,
\end{aligned}
\\
\label{eq:38}
&
\begin{aligned}
\tilde{\tilde{H}}_{SB}(t, \tau)=
\sum_{i}
\tilde{\tilde{C}}_i^{\dagger}(\tau) \tilde{\tilde{d}}_i(t, \tau)
+
\tilde{\tilde{d}}_i^{\dagger}(t, \tau) 
\tilde{\tilde{C}}_i(\tau)~~~~~~~~~~~~
\end{aligned}
\end{eqnarray}
where $\hat{\rho}_S(t)$ and $\hat{\rho}_B$ are the System and bath many-body density operators. 
$\operatorname{Tr}_B$ indicates tracing over the bath degrees of freedom.
In the, 
Eq. (\ref{eq:38}), $\tilde{\tilde{C}}_i(\tau)=\sum_{l k} V_{l k,i}^{\ast} \hat{c}_{l k} e^{i \, \epsilon_{lk}  \tau}$ and it determines how the bath part of the $\hat{H}_{SB}$ evolve in the bath's time scale, $\tau$.
On the other hand, $\tilde{\tilde{d}}_i(t, \tau) =\hat{U}_S(t, t-\tau) \hat{d}_i \hat{U}_S^{\dagger}(t, t-\tau)$ describes the evolution of the System part of the $\hat{H}_{SB}$ in which the time evolution operator is $\hat{U}_S(t, t-\tau)=\mathcal{T} exp\big({-i \int_{t-\tau}^t \hat{H}_S(s) d s}\big)$  ($\mathcal{T}$ refers to the time-ordering operator). Same definition holds for $\tilde{\tilde{d}}^{\,{\dagger}}_i(t, \tau)$. It was Shirley who first formalized the time evolution operator for the time-periodic Hamiltonian. The time evolution operator can be expressed as
\begin{eqnarray}
\label{eq:39}
\begin{aligned}
\hat{U}_S(t, t_0)=
\sum_n\langle n|e^{-i \hat{H}_S^F\left(t-t_0\right)}| 0\rangle e^{i n \omega t},
\end{aligned}
\end{eqnarray}
where $| n \rangle$ ($| 0 \rangle$) refers to the $n$th ($0$th) state in the Fourier basis set. Here, the Floquet many-body Hamiltonian defines identical to the one-body counterpart as $\hat{H}_S^F=\sum_{n,m}\hat{H}_{S\, m n}-\hat{I} (m \omega) \delta_{mn}$ in which $\hat{H}_{S\,m n}$ defines similar to Eq. (\ref{eq:12}).  
In general, disassembling the double commutator in Eq. (\ref{eq:37}) results in eight major integrands. In addition, having two parts in Eq. (\ref{eq:38}) results in a total of sixteen integrands. From these, only eight integrands should be considered (terms in which one of the operators multiples to the conjugate transpose of the other). For example, the first and second non-vanishing terms (\textit{Dissipators}) can be given as
\begin{eqnarray}
\label{eq:40}
\begin{aligned}
&
\sum_{i,j} 
\int_0^{\infty} 
\operatorname{Tr}_B
\Big(
\big(
\hat{C}_{i}^\dagger \hat{d}_i+
\hat{d}_i^\dagger \hat{C}_{i}
\big)
\big(
\tilde{\tilde{C}}_j^{\dagger}(\tau) \tilde{\tilde{d}}_j(t, \tau)
+
\tilde{\tilde{d}}_j^{\dagger}(t, \tau)
\\&
\tilde{\tilde{C}}_j(\tau)
\big)
\,
\hat{\rho}_S(t) \otimes \hat{\rho}_B
\Big)
 d \tau=
\sum_{i,j} 
\int_0^{\infty} 
\operatorname{Tr}_B
\Big(
\hat{C}_{i}^\dagger \hat{d}_i
\tilde{\tilde{d}}_j^{\dagger}(t, \tau)~~~~~~
\\&
\tilde{\tilde{C}}_j(\tau)
\hat{\rho}_S(t) \otimes \hat{\rho}_B
+
\hat{d}_i^\dagger \hat{C}_{i}
\tilde{\tilde{C}}_j^{\dagger}(\tau) 
\tilde{\tilde{d}}_j(t, \tau)\hat{\rho}_S(t) \otimes \hat{\rho}_B
\Big)d \tau.
\end{aligned}
\end{eqnarray}
For the sake of brevity, we only focus on the first Dissipator which can be further simplified to 
\begin{eqnarray}
\label{eq:41}
\begin{aligned}
&
\sum_{i,j} 
\int_0^{\infty} 
\operatorname{Tr}_B
\Big(
\hat{C}_{i}^\dagger \tilde{\tilde{C}}_j(\tau)
\hat{\rho}_B
\otimes
\hat{d}_i
\tilde{\tilde{d}}_j^{\dagger}(t, \tau)
\hat{\rho}_S(t)  
\Big)d \tau= \\
&
\sum_{i,j,kl} 
\int_0^{\infty} 
\big(
V_{l k,i} V_{l k,j}^{\ast}  
e^{i \,\epsilon_{lk} \tau}
f(\epsilon_{lk}, \mu_l)
\hat{d}_i
\tilde{\tilde{d}}_j^{\dagger}(t, \tau)
\hat{\rho}_S(t)  
\big)
d \tau.
\end{aligned}
\end{eqnarray}
Note that 
$\operatorname{Tr}_B
\Big(\hat{c}_{lk}^{\dagger}
\hat{c}_{l^\prime k^\prime} \hat{\rho}_B(\mu)
\Big)=f(\epsilon_{lk}, \mu_l)\delta_{k,k^\prime}\delta_{l,l^\prime}$. Until now, the derivation procedure is not far from the non-Floquet QME. However, complexity in the time dependency of the $\tilde{\tilde{d}}_j^{\,{\dagger}}(t, \tau)$ requires Shirley's Floquet Formalism, Eq. (\ref{eq:39}), as
\begin{eqnarray}
\label{eq:42}
\begin{aligned}
\tilde{\tilde{d}}_j^{\,\dagger}(t, \tau)=
\sum_{n, m}
\langle n|
e^{-i \hat{H}^F_S\tau} 
|0\rangle 
\hat{d}_j^{\,\dagger}
\langle 0|
e^{i \hat{H}^F_S\tau}
| m\rangle 
e^{i(n-m) \omega t}.~~~~
\end{aligned}
\end{eqnarray}
To proceed, we shall use the eigenbasis of the Floquet System Hamiltonian $\hat{Y}^{\dagger} \hat{H}^F_S \hat{Y}=\hat{\Lambda}^F$ in which $\hat{Y}$ is the rotating operator. Then, the above relation can be simplified to
\begin{eqnarray}
\label{eq:43}
\begin{aligned}
\tilde{\tilde{d}}_j^{\,\dagger}(t, \tau)=
\sum_{n, m}
\langle n|
\hat{Y}
e^{-i \hat{\Lambda}^F\tau}  
\hat{\mathcal{D}}_{j}^{{o}\,\dagger} 
e^{i \hat{\Lambda}^F\tau}
\hat{Y}^{\dagger} 
| m\rangle 
e^{i(n-m) \omega t},~~~~
\end{aligned}
\end{eqnarray}
where $\hat{\mathcal{D}}_{j}^{{o}\,\dagger}=
\hat{Y}^{\dagger} 
\big(|0\rangle 
\hat{d}_j^{\,\dagger}
\langle 0|\big)
\hat{Y}$ and we used the property $\hat{Y}^{\dagger}\hat{Y}=\hat{I}$ multiple times.
Next, we should consider a basis set $\{|a\rangle \}$ for the Hilbert space which allows us to present elements of density operator as $\rho_{ab}(t)= \langle a |\hat{\rho}(t)|b\rangle$ and turn operators in Eq. (\ref{eq:43}) to matrices. In particular, it allows us to work with the (hybrid) Floquet basis, $\{| \gamma \rangle \}$ such that we can simplify the $e^{-i\hat{\Lambda}^F\tau}  
\hat{\mathcal{D}}_{j}^{{o}\,\dagger} 
e^{i \hat{\Lambda}^F\tau}$. A matrix element of this term is given by 
\begin{eqnarray}
\label{eq:44}
\begin{aligned}
\Big(
e^{-i \hat{\Lambda}^F\tau}  
\hat{\mathcal{D}}_{j}^{{o}\,\dagger} 
e^{i \hat{\Lambda}^F\tau}
\Big)_{\gamma,\nu}=
e^{-i \Omega_{\gamma \nu} \tau}  
\big(
\hat{\mathcal{D}}_{j}^{{o}\,\dagger}
\big)_{\gamma,\nu},
\end{aligned}
\end{eqnarray}
where $\Omega_{\gamma \nu}=(\mathcal{E}_{\gamma} -\mathcal{E}_{\nu})$. Clear separation of the two time scales ($t$ and $\tau$) allows us to perform the time integration over $\tau$ as  
\begin{eqnarray}
\label{eq:45}
\begin{aligned}
&\int_0^{\infty}   
e^{i \,(\epsilon_{lk}-\Omega_{\gamma \nu}) \tau}
d \tau=
\pi \delta(\epsilon_{lk}+\Omega_{\gamma \nu})
-iP
\big( 
\frac {1}{\epsilon_{lk}+\Omega_{\gamma \nu} }
\big).~~
\end{aligned}
\end{eqnarray}
Neglecting the Cauchy's principle, $P$, and employing the energy-independent wide-band approximation, $\Gamma^l_{ij}=2\pi \sum_kV_{l k,i} V_{l k,j}^{\ast}  $ results in the following matrix expression for the first Dissipators
\begin{eqnarray}
\label{eq:46}
\begin{aligned}
&
\sum_{i,j,l} 
\frac {\Gamma^l_{ij} }{2}
{d}_i
\sum_{n, m}
\langle n|
{Y} 
f_l(-\Omega) 
\circ
{\mathcal{D}}_{j}^{{o}\,\dagger} 
{Y}^{\dagger} 
| m\rangle 
e^{i(n-m) \omega t}
{\rho}_S(t)  
.~~
\end{aligned}
\end{eqnarray}
where $\circ$ refers to the Hadamard product. We should repeat this procedure for the other seven non-vanishing Dissipators. A more compact matrix form of the Floquet QME in the Hilbert space can then be given as
\begin{eqnarray}
\label{eq:47}
\begin{aligned}
\begin{gathered}
\frac{\partial \rho_S(t)}{\partial t}=
-i
\left[
{H}_S(t), {\rho}_S(t)
\right]
-\sum_{i,j,l}
\frac{\Gamma^l_{ij}}{2}
\Big(d_i\widetilde{d}_{jl}^{\, \dagger}(t)\rho_S(t)+~~~
\\
{d}_i^{\, \dagger} \overline{d}_{jl}(t) \rho_S(t)-d_i \rho_S(t) \widetilde{d}_{jl}^{\, \dagger}(t)
-{d}_i^{\dagger} \rho_S(t) \overline{d}_{jl}(t)
\Big)+\text { h.c. },
\end{gathered}
\end{aligned}
\end{eqnarray}
where $\widetilde{d}_{jl}(t)$ and $\overline{d}_{jl}(t)$ defined as
\begin{eqnarray}
\label{eq:48}
\widetilde{d}_{jl}^{\, \dagger}(t)
&=& 
\sum_{n, m}
\langle n|
{Y} 
\,
f_l(-\Omega) 
\circ
{\mathcal{D}}_{j}^{{o}\,\dagger}
\,
{Y}^{\dagger} 
| m\rangle 
e^{i(n-m) \omega t},~~~~
\\
\label{eq:49}
\overline{d}_{jl}(t)
&=&\sum_{n, m}
\langle n|
{Y} 
\,
f_l(\Omega) 
\circ
{\mathcal{D}}_{j}^{{o}} 
\,
{Y}^{\dagger} 
| m\rangle 
e^{i(n-m) \omega t}.
\end{eqnarray}
We stress that within the Hilbert-Space FQME, $\widetilde{d}_{jl}(t)$ and $\overline{d}_{jl}(t)$ (and their hermitian conjugates) are time-dependent matrices. Finally, we can simplify the notation of Eq. (\ref{eq:47}) as ${\partial \rho_S(t)}/{\partial t}=-i
\left[
{H}_S(t), {\rho}_S(t)
\right]+\sum_{l} \mathcal{L}_l^{HS}(t)\rho_S(t)$. 
\subsubsection{Floquet-Space FQME}
LvN equation is the starting point for varieties of QMEs (e.g. the Redfield QME). For a periodic Hamiltonian, the LvN equation can be transformed into the Floquet LvN equation which is very general and has a long history in the analysis of nuclear magnetic resonance signals (closed systems) \cite{ivanov2021floquet}. The Floquet LvN equation is given by
\begin{eqnarray}
\label{eq:50}
\frac{\partial \hat{\rho}^F(t)}{\partial t}
&=&
-i
\left[
\hat{H}^F, \hat{\rho}^F(t)
\right]
\\
\label{eq:51}
\hat{H}^F
&=&
\sum_{n}
\widehat{L}_n \otimes \hat{H}^{(n)} 
+\widehat{N} \otimes \hat{\mathbf{1}},
\\
\label{eq:52}
\hat{H}^{(n)}
&=&
\frac{1}{T} \int_0^T dt\,  e^{i n \omega t} \hat{H}(t) \\
\label{eq:53}
\hat{\rho}^F(t)
&=&
\sum_{n}
\widehat{L}_n \otimes \hat{\rho}^{(n)}(t).
\end{eqnarray}
Here, $n$ is an integer in the space $[-N, N]$, $\widehat{L}_{n}$ is the Ladder operator of the order $n$, $\widehat{N}$ is the Number operator in the Fourier space, and $\hat{\mathbf{1}}$ is the identity operator in the Hilbert space. In the matrix form, $\widehat{L}_{n}$ represents a $N\times N$ off-diagonal matrix with ones in its off-diagonal and $\widehat{N}$ represents a diagonal matrix with the integer vector $\{-N,..., N\}$ placed on it's diagonal. 
The major benefit of the Floquet LvN equation is that $\hat{H}(t)$ turns into a time-independent Hamiltonian $\hat{H}^F$.  
Note that the operator coefficient $\hat{\rho}^{(n)}(t)$ is time dependent whereas $\hat{H}^{(n)}$ is not. The definition of $\hat{H}^{(n)}$ is consistent with Eq. (\ref{eq:12}) and hence the $\hat{H}^F$ given in Eq. (\ref{eq:51}) is effectively similar to those discussed before.
 Assuming that $\hat{H}(t)$ is the total time-periodic Hamiltonian, one can partition the $\hat{H}^F$ as
\begin{eqnarray}
\label{eq:54}
\hat{H}^F
&=&
\hat{H}^F_S \otimes \hat{I}_B+
\hat{I}^F_S \otimes \hat{H}_B+
\hat{H}^F_{SB}, \\
\label{eq:55}
\hat{H}_{SB}^F 
&=&
\sum_{i} 
\hat{C}_{i}^\dagger \hat{\mathcal{D}}^{}_i +\mathrm{H.c.}
\end{eqnarray}
where $\hat{I}^F_S=\widehat{L}_0 \otimes \hat{I}_S$ and $\hat{\mathcal{D}}^{}_i= \widehat{L}_0 \otimes \hat{d}_i$. Then, we can define the following rotation protocols   
\begin{eqnarray}
\label{eq:56}
\tilde{H}^F_{BS}(t)
&=&
e^{i \hat{H}_B t} 
e^{i \hat{H}^F_S t} 
\hat{H}^F_{BS} 
e^{-i \hat{H}_B t} 
e^{-i \hat{H}^F_S t}, \\
\label{eq:57}
\tilde{\rho}(t)
&=&
e^{i \hat{H}_B t} 
e^{i  \hat{H}^F_S t} 
\hat{\rho}(t) 
e^{-i \hat{H}_B t} 
e^{-i \hat{H}^F_S t}.
\end{eqnarray}
The above rotations follow the transformation to the interaction picture in the context of conventional QME but, here within this version of Floquet QME, we have used $\hat{H}^F_{S/SB}$. Next, we can take the (initial factorization) assumption $\tilde{\rho}^{F}(t)=\tilde{\rho}^F_S(t) \otimes \tilde{\rho}_B$ and proceed just as the conventional QME. This eventually leads to the following Floquet Markovian relation in the Schr\"odinger frame
\begin{eqnarray}
\label{eq:58}
\begin{aligned}
\frac{\partial \hat{\rho}_S^F(t)}{\partial t}
&=-i\left[\hat{H}_S^F, \hat{\rho}_S^F(t)\right]-\int_0^{\infty} \operatorname{Tr}_B
\\
&
\quad
\left[\hat{H}^F_{SB},
\left[\tilde{\tilde{H}}^F_{S B}(\tau), \hat{\rho}_S^F(t) \otimes \hat{\rho}_B\right]\right] d \tau,
\end{aligned}
\end{eqnarray}
where $\tilde{\tilde{H}}^F_{S B}(\tau)=e^{-i \hat{H}_B \tau} e^{-i \hat{H}_S^F \tau} \hat{H}^F_{S B} e^{i \hat{H}_B \tau} e^{i \hat{H}_S^F \tau}$. 
Because $\hat{\rho}_S^F(t)$ is in the Floquet space, we can employ Floquet basis, $\{| \gamma \rangle \}$, to present elements of the reduced Floquet density operator (Floquet Hamiltonian) as $\hat{\rho}^F_{S\, \gamma \nu}= \langle \gamma |\hat{\rho}_S^F(t)| \nu \rangle$
($H^F_{S\, \gamma \nu}(t)= \langle \gamma | \hat{H}_S^F | \nu \rangle$).
After taking the integrations over $\tau$ and employing Wide-band approximation, we arrive at the following matrix expression for the dynamic of the Floquet density matrix 
\begin{eqnarray}
\label{eq:59}
\begin{aligned}
\begin{gathered}
\frac{\partial \rho^F_S(t)}{\partial t}=
-i
\left[
{H}^F_S, {\rho}^F_S(t)
\right]
-
\sum_{i,j,l} 
\frac{\Gamma_{ij}^{l}}{2}
\Big(\mathcal{D}_i\widetilde{\mathcal{D}}_{jl}^{\, \dagger}\rho^F_S(t)+~~~
\\
\mathcal{D}_i^{\, \dagger} \overline{\mathcal{D}}_{jl}\rho^F_S(t)
-\mathcal{D}_i \rho^F_S(t) \widetilde{\mathcal{D}}_{jl}^{\, \dagger}
-\mathcal{D}_i^{\dagger} \rho^F_S(t) \overline{\mathcal{D}}_{jl}
\Big)+\text { h.c. },
\end{gathered}
\end{aligned}
\end{eqnarray}
where $
\widetilde{\mathcal{D}}_{jl}^{\, \dagger} = {Y} \,f_l(-\Omega) \circ{\mathcal{D}}_{j}^{{}\,\dagger}\,{Y}^{\dagger}$
and  $\overline{\mathcal{D}}_{jl}={Y} \,f_l(\Omega) \circ{\mathcal{D}}_{j}^{} \,{Y}^{\dagger}$. 
The Floquet-Space FQME is thus easier to be implemented as compared to the Hilbert-Space FQME. The disadvantage of Floquet-Space FQME is that it requires more memory since all its components are in the Floquet space. Finally, we can simplify the notation of Eq. (\ref{eq:59})  as ${\partial \rho^F_S(t)}/{\partial t}=-i
\left[
{H}^F_S, {\rho}^F_S(t)
\right]+ \sum_ {l}\mathcal{L}_l^{FS}\rho^F_S(t)$. Note that $\mathcal{L}_l^{FS}$ is time-independent. 
\subsection{Observable in Floquet QME }
Within the Hilbert-Space FQME, it is straightforward to obtain expectation values of the particle number, $\langle \hat{n} \rangle$ and current passing through the contact $l$, $\langle {\hat{J}_l}\rangle$, \cite{harbola2006quantum} as
\begin{eqnarray}
\label{eq:60}
\langle \hat{n} \rangle(t)
&=&
Tr\big(\hat{n} \hat{\rho}(t)\big), \\
\label{eq:61}
\langle \hat{J}_l \rangle(t)
&=&
Tr\big(\hat{n} \mathcal{L}_l^{HS}(t) \hat{\rho}(t)\big).
\end{eqnarray}
To evaluate the expectation values of an observable in the Floquet-Space FQME, we need to project the operators from the Floquet space to the Hilbert space \cite{ivanov2021floquet} as
\begin{eqnarray}
\label{eq:62}
\langle \hat{n} \rangle(t)
&=&
\sum_m
Tr\big(
\langle m|
\hat{n}^F \hat{\rho}^F(t)
|0\rangle e^{i m \omega t}
\big),
\\
\label{eq:63}
\langle \hat{J}_l \rangle(t)
&=&
\sum_m
Tr\big(
\langle m|
\hat{n}^F
\mathcal{L}_l^{FS} 
\hat{\rho}^F(t)
|0\rangle e^{i m \omega t}
\big).
\end{eqnarray}
To evaluate $\overline{n}$ and $\overline{J}_l$ in both Floquet QME methods, we must evaluate the time-average of observables, $\hat{O}(t)$, numerically by using $\overline{O}=1/T \int_0^T \langle \hat{O} \rangle(t) dt$. 
\section{Numerical Results}
To demonstrate the implications of light-matter interactions on the transport characteristics of an open quantum system, and to examine 
{\color{black} agreement} between the discussed methods (V-like Floquet NEGF, M-like Floquet NEGF, Hilbert-Space FQME, and Floquet-Space FQME), we have first taken a spinless two-level quantum dot placed between two metallic contacts. 
\begin{figure}
	\begin{center}
		\includegraphics[width=6.0cm]{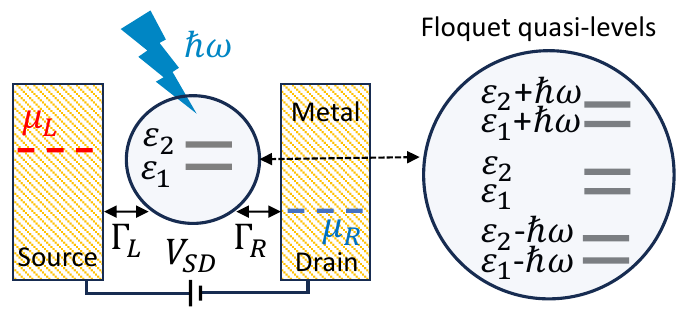}
	\end{center}
	\caption{\label{fig:1} (a) Schematics of a quantum transport system comprised of a two-level System under light illumination. The two-level system with monochromatic light can roughly be thought of as multiple quasi-levels.}
\end{figure}
This two-level dot then interacts with a monochromatic light. The one-body Hamiltonian of a such quantum dot can be given by
\begin{eqnarray}
\label{eq:64}
[h](t)= \left[
\begin{array}{cc}
\varepsilon_{1} & A \cos (\omega t) \\
A \cos (\omega t) & \varepsilon_{2}
\end{array}
\right],
\end{eqnarray}
where $\omega$ is the frequency of light and $A$ represents the light intensity. Off-diagonal elements on the above Hamiltonian represent the dipole approximation in the context of light-matter interaction (Rabi model in quantum optics). 
This Hamiltonian is connected to the left (right) contacts with small $\Gamma_{L(R)}$ (weak coupling regime). The schematic of such a configuration is given in Fig.~\ref{fig:1}.
\subsection{Results for Floquet NEGF methods}
As the first example, we calculate the $\overline{n}$ and $\overline{J}_L$ (current passing through the left contact) using Eqs. (\ref{eq:25})-(\ref{eq:28}) for a range of external bias, $V_{SD}=\mu_R-\mu_L$ and multiple light frequencies. 
{\color{black} Except the $N$, other parameters will be in eV units and we will omit the unit for simplicity.}
Hereafter, we will fix the right contact (Drain) at $\mu_R=-0.4$ and sweep over $\mu_L$. 
{\color{black} 
Note that in Floquet NEGF methods, care should be taken in choosing the energy steps.}
In the absence of the {\color{black}Floquet driving (light)}, and when off-diagonals of $[h]$ are zero, there should be only two plateaus on {\color{black}
$\overline{n}$-$\mu_L$ and  $\overline{J}$-$\mu_L$ curves} 
such that the onset of these two plateaus are at $\varepsilon_{1}$ and $\varepsilon_{2}$.  
When the $A$ is large enough and the frequency of light is tuned to the difference between two levels, $\hbar\omega=\varepsilon_{2}-\varepsilon_{1}$, then maximum alteration of the quantized {\color{black} steps} will happen such that the number of plateaus is doubled. This has been shown in Figs.~\ref{fig:2}(a) and \ref{fig:2}(b). The extra quantized plateaus can be the hallmark of an electron-photon hybrid state (Floquet states). 
\begin{figure}
	\begin{center}
		\includegraphics[width=5.0cm]{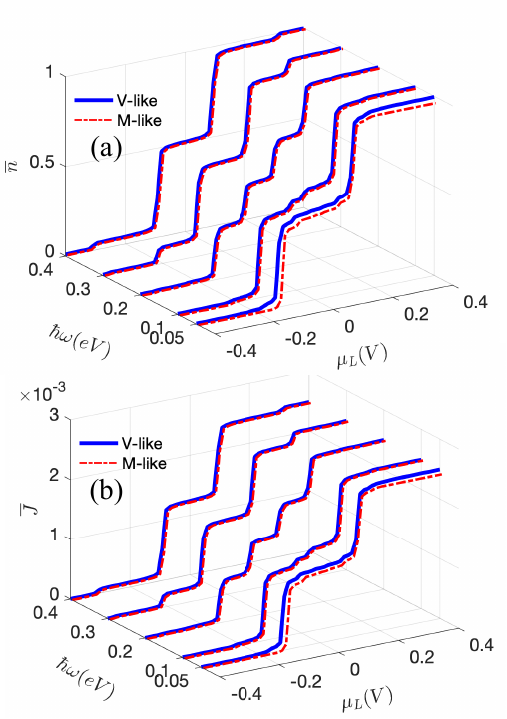}
	\end{center}
  \captionsetup{labelfont={color=black},font={color=black}}
	\caption{\label{fig:2} Floquet NEGF: (a) One-period time-average of the number operator, $\overline{n}$, for multiple driving frequencies. Parameters are: $\varepsilon_{1}=-0.1$, $\varepsilon_{2}=0.1$, $A=0.1$, $N=3$, $kT=0.0036~(4.2~K)$, $\Gamma_{L}=\Gamma_{R}=0.0025$, $\mu_R=-0.4$.}
\end{figure}
\begin{figure}
	\begin{center}
		\includegraphics[width=5.0cm]{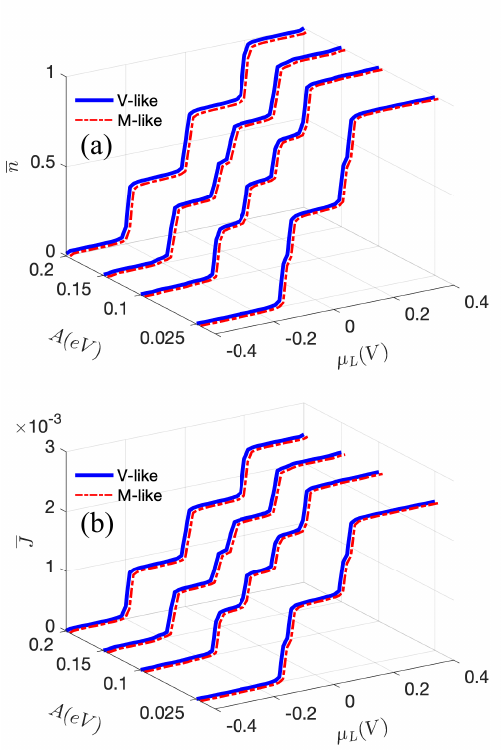}
	\end{center}
	\caption{\label{fig:3} Floquet NEGF: (a) One-period time-average of the number operator, $\overline{n}$, and (b) one-period time-average of the left contact's current operator, $\overline{J}_L$, for multiple driving amplitudes at {\color{black}$\hbar \omega=0.2$}. All other parameters are the same as Fig.~\ref{fig:2}.}
\end{figure}
In extreme off-tuning situations (associated with $\hbar\omega=0.05$ and $\hbar\omega=0.4$), the transport characteristics should be almost identical to the non-Floquet transport characteristics. We can see it from the first and last curves in Figs.~\ref{fig:2}(a) and \ref{fig:2}(b). 
\FloatBarrier 
\onecolumngrid 
\begin{figure*}[!t]
	\centering
	 \includegraphics[width=14.5cm]{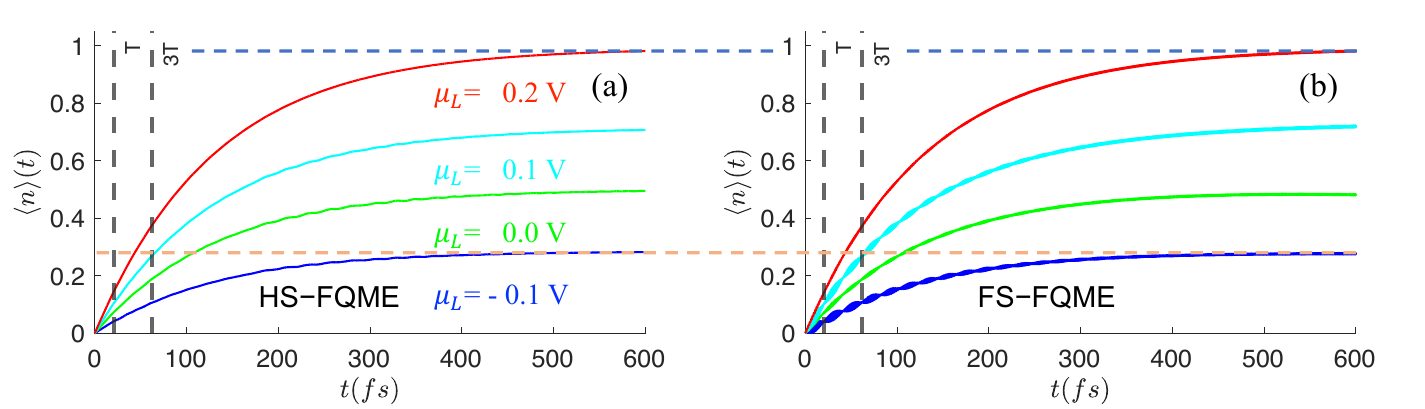}
 \captionsetup{labelfont={color=black},font={color=black}}
	\caption{\label{fig:4} (color online) Floquet QME: (a) Time-dependent expectation value of the number operator evaluated by the Hilbert space Floquet QME (HS$-$FQME) for a few of external biases. (b) Same as (a) evaluated by the Floquet space Floquet QME (FS$-$FQME) at $\hbar \omega=0.2$. Other parameters are the same as Fig.~\ref{fig:2}.}
\end{figure*}
\twocolumngrid 
At low frequencies, the M-like Floquet NEGF does not provide accurate results with only $N=3$ 
{\color{black}and $N$ has to be increased.} Hence, V-like Floquet NEGF delivers better numerical performance in low frequencies.
Next, we fixed the driving frequency at 
{\color{black} $\hbar \omega=0.2$}
and calculated the two observables for a range of external bias, $\mu_L$, and multiple driving amplitudes, $A$. Results are plotted in Figs.~\ref{fig:3}(a) and \ref{fig:3}(b). Here, we can see, that {\color{black} the driving amplitude $A$, shifts the onset of plateaus symmetrically around the center of energy between $\varepsilon_{1}$ and $\varepsilon_{2}$}. Comparing evaluations in Fig.~\ref{fig:2} and Fig.~\ref{fig:3}, with respect to increasing $\hbar \omega$ (light frequency) and $A$ (light intensity), we can see that increasing $A$ splits the heights equally into two steps (see small dents in the curves of $A=0.025$ in Fig.~\ref{fig:3} ) and shift the onsets of newly born steps, whereas increasing $\hbar \omega$ modifies the heights linearly. 
\subsection{Results for Floquet QME methods}
We now use the spinless many-body operators $\hat{d}/\hat{d}^\dagger$ to make the time-dependent many-body Hamiltonian and its Floquet counterpart. Then, by identifying the many-body basis set, an initial state (initial many-body density matrix), and the Fourier space (hence the many-body Floquet space), we will be able to perform dynamical solutions to Eqs. (\ref{eq:47})  and (\ref{eq:59}). 
Afterward, we can evaluate the time averages of the observables, Eqs. (\ref{eq:60})-(\ref{eq:63}).
Note that, we have used the Runge-Kutta 4th-order method to solve Floquet QMEs. The time step has to be adjusted carefully. The rest of the parameters are the same as what is given in Fig.~\ref{fig:2}. Here, we consider the perfect tuning case, $\hbar\omega=0.2$, with the amplitude $A=0.1$. The dynamic of the total population on the dot, $\langle \hat{n} \rangle(t)$, calculated by Hilbert space Floquet QME (HS$-$FQME) and Floquet space Floquet QME (FS$-$FQME) methods are plotted in Figs.~\ref{fig:4}(a) and \ref{fig:4}(b). For the initial state, we have considered both levels to be empty at $t=0$. Interestingly, we can see a pronounced oscillation in the $\langle \hat{n} \rangle(t)$ of the FS$-$FQME [Fig.~\ref{fig:4}(b)] in the short-time limit, whereas the dynamics of HS$-$FQME [Fig.~\ref{fig:4}(a)], does not feature such oscillation in the early times. Both Floquet QME methods converge to a similar steady state on the long-time limit (marked by dashed lines) resulting in similar time averages for the expectation values (see Fig.~\ref{fig:5R}). 
\begin{figure}
	\begin{center}
		\includegraphics[width=6.0cm]{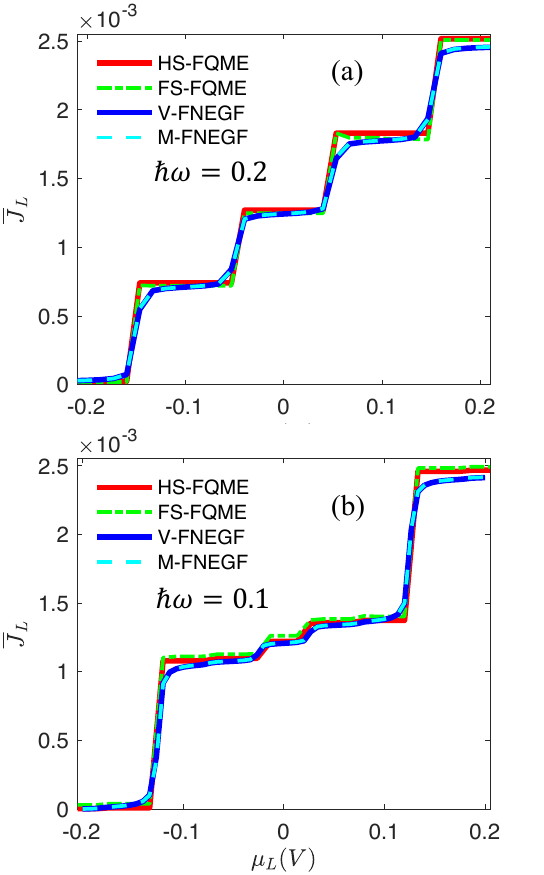}
	\end{center} 
 \captionsetup{labelfont={color=black},font={color=black}}
 \caption{\label{fig:5R} (color online) (a) One-period time-average of the left contact's current, $\overline{J}_L$, at $\hbar \omega=0.2$ for HS$-$QME, FS$-$QME, V$-$FNEGF, and M$-$FNEGF method. HS$-$QME and FS$-$QME are corresponding with Figs.~\ref{fig:4}(a) and \ref{fig:4}(b), respectively, while V$-$FNEGF and M$-$FNEGF are corresponding with Fig.~\ref{fig:2}. (b) Same as (a) for $\hbar \omega=0.1$.}
\end{figure}
 
In Figs.~\ref{fig:5R}(a) and \ref{fig:5R}(b), we compare the average current calculated by HS$-$QME and FS$-$QME with each other as well as with the V-like Floquet NEGF (V$-$FNEGF) and M-like Floquet NEGF (M$-$FNEGF) methods for two different driving frequencies. For the small $\Gamma$, we can see a quantitative agreement between the four methods. 
This agreement is one of the main results of this
work. Edge of current plateaus are smooth in Floquet NEGF methods which indicates Floquet NEGF methods can capture broadening effects whereas, Floquet QME methods maintain their disability to incorporate broadening effects. Note that, we have found both Floquet QMEs require much smaller time steps as compared to the non-Floquet QME. However, in comparison, the HS$-$FQME is less sensitive to the smallness of the time step than the FS$-$FQME. Our extensive numerical examinations show also that FS$-$FQME requires a larger $N$ for larger $\Gamma$s. Nonetheless, the four methods agree with each other by choosing the appropriate $N$ and the time/energy step. 
\subsection{Results for Floquet driven interacting system}
{\color{black} We now focus on the interplay between the Floquet driving and the strength of electron-electron interactions {\color{red} first by} using the HS$-$FQME method. 
In Fig.~\ref{fig:6R}(a), we show the current characteristics of the two-level spin system when subjected to a Floquet driving in the absence and the presence of multiple electron-electron interactions.
\begin{figure}
	\begin{center}
		\includegraphics[width=6.0cm]{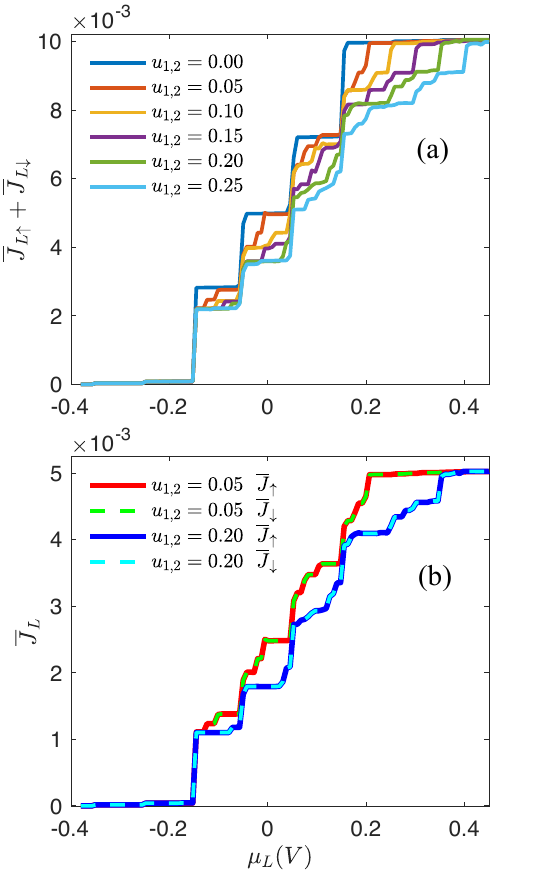}
	\end{center}
 \captionsetup{labelfont={color=black},font={color=black}}
	\caption{\label{fig:6R} (color online) Floquet QME for the interacting case: (a) One-period time-average of the total left contact's current, $\overline{J}_{L\uparrow}+\overline{J}_{L\downarrow}$, at $\hbar \omega=0.2$ in the absence and the presence of multiple electron-electron interactions. Here, $\Gamma =0.005$, and other parameters are similar to Fig.~\ref{fig:2}. (b) The spin-resolved of (a) only for two values of the electron-electron interaction.} 
\end{figure}
Here, we have considered the dipole approximation to drive both spins in the same way. For the non-interacting case, $u_{1,2}=0$, having two spins, obviously, results in the degeneracy factor $2$ on the observed plateaus. We can see there is an interesting interplay between the interaction and the Floquet driving interacting cases.
Here, with the real value $A$, we would not have spin polarization, as can be seen from Fig.~\ref{fig:6R}(b), where we have shown spin-resolved time average current only for two values of the electron-electron interaction. 
Floquet driving produces finer quantized plateaus (in the averaged current) in the presence of smaller electron-electron interactions. Once $u_{1,2}>\hbar\omega$, the interacting effect dominates such that the two lowest plateaus become flat again, see the curve of $u_{1,2}=0.25$ in Fig.~\ref{fig:6R}(a). 
{\color{black}
Next, in Fig.~\ref{Fig6_7R2_2} (a), we have explored the degree of agreement between the Floquet HS$-$QME (FQME) and Floquet interacting NEGF (FINEGF) by comparing the averaged current for a strongly interacting regime ($u_i > \varepsilon_{2}-\varepsilon_{1}$) with $u_{1,2}=0.3$. 
Here,  FINEGF refers to a combination of Eqs. (\ref{eq:33})-(\ref{eq:36}) and Eq. (\ref{eq:28}), whereas FQME refers to Eqs. (\ref{eq:47})-(\ref{eq:49}) and Eq. (\ref{eq:61}).
For reference, the current characteristics of the non-driven interacting QME \cite{dou2016many} and interacting NEGF (INEGF) are also plotted against each other in Fig.~\ref{Fig6_7R2_2} (b). Fig.~\ref{Fig6_7R2_2} (a) illustrates how driving substantially alters the transport characteristics even in the presence of strong electron-electron interaction. Importantly, in the presence of relatively strong electron-electron interaction, the onset of driven-induced quantized plateaus is consistent between FQME and FINEGF approaches. However, the simplest Floquet NEGF for an interacting system fails to predict the correct height for each current step. Fig.~\ref{Fig6_7R2_2} (a)  demonstrates that unless the height of the current is the focus of a numerical simulation, the ansatz given by Eqs. (\ref{eq:36})-(\ref{eq:39}) delivers a satisfactory outcome, despite the fact that it is not a self-consistent method.
}
\begin{figure}
	\begin{center}
	\includegraphics[width=6.0cm]{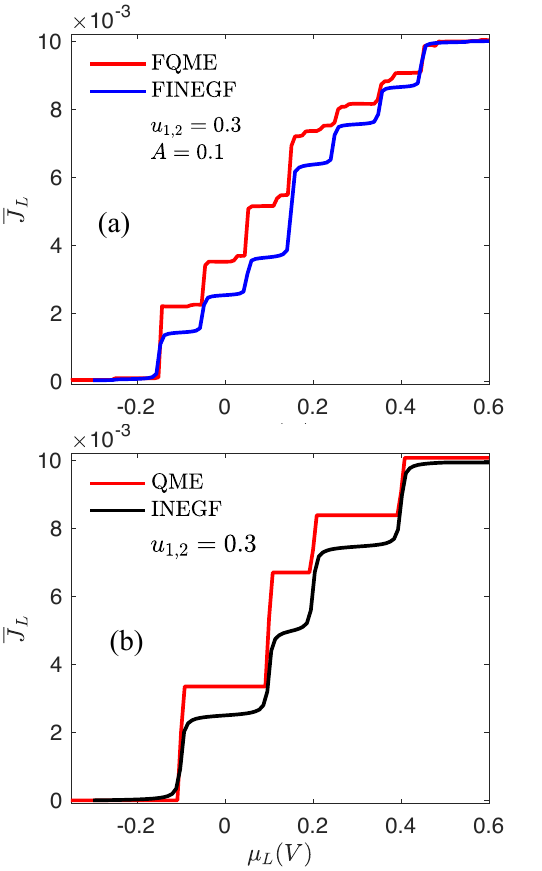}
	\end{center}
 \captionsetup{labelfont={color=black},font={color=black}}
	\caption{\label{Fig6_7R2_2} (color online) Comparison between results of QME and NEGF for an interacting case: (a) Average current on the left contact evaluated by Hilbert-Space Floquet QME (FQME) and Floquet (M-like) NEGF (FINEGF). (b) Same as (a) in the absence of driving. Here, $\Gamma =0.005$, $\hbar \omega=0.2$  and other parameters are similar to Fig.~\ref{fig:2}.} 
\end{figure}
}
\subsection{Interacting system with a complex Floquet driving}
Finally, we consider a case in which the Floquet driving is a complex time-dependent function.
In Fig.~\ref{fig:7R}, we have plotted the spin-resolved time average current for two Hamiltonian models in the presence and absence of the electron-electron interaction.  
In model 1, the one-body Hamiltonian for both spin-up and spin-down, $[h]_{\uparrow,\downarrow}(t)$, are identical and are given by: 
\begin{eqnarray}
\label{eq:67R}
\left[
\begin{array}{cc}
\varepsilon_{1} & A(cos(\omega t)+i~sin(\omega t)) \\
A(cos(\omega t)-i~sin(\omega t)) & \varepsilon_{2}
\end{array}
\right].~~~~
\end{eqnarray}
In addition, we have considered $A=0.1/\sqrt{2}$ (to keep the total power of the incident light similar to the previous examples). However, in model 2, we take $[h]_{\downarrow}(t)=[h]_{\uparrow}(t)^*$. These complex drivings can represent different (circular) polarizations of the external light.
\begin{figure}
	\begin{center}
		\includegraphics[width=8.6cm]{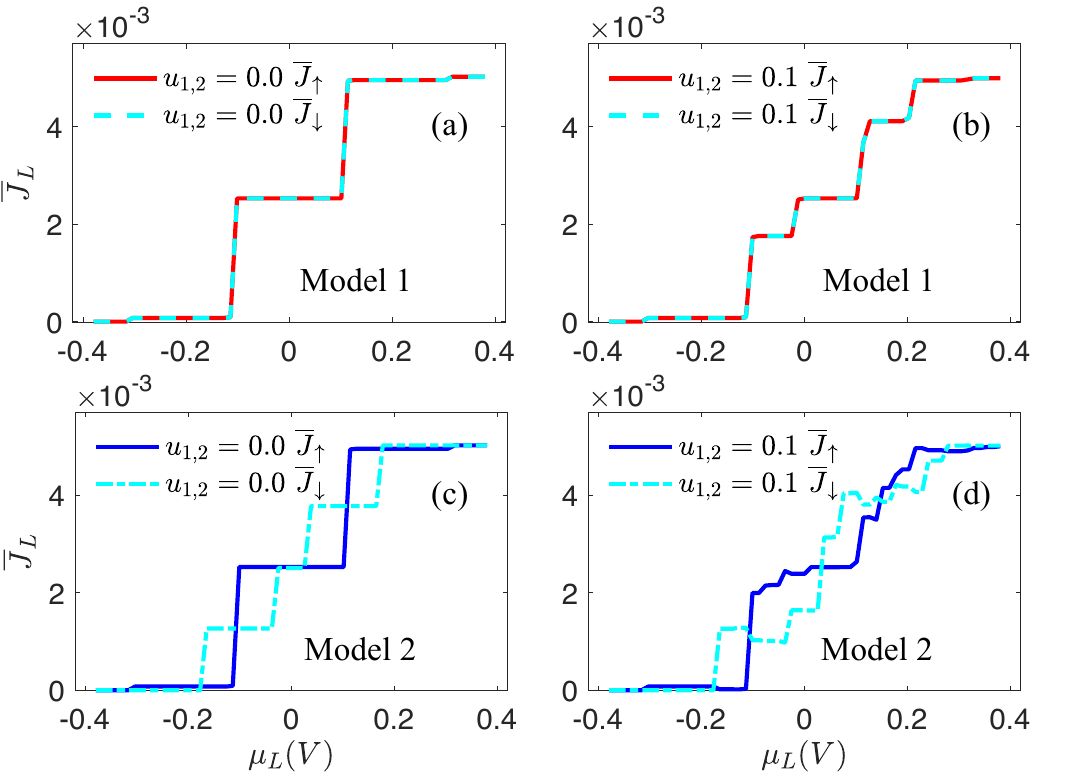}
	\end{center}
 \captionsetup{labelfont={color=black},font={color=black}}
	\caption{\label{fig:7R} (color online) One-period time average of the left contact spin-up, $\overline{J}_{\uparrow}$, and spin-down, $\overline{J}_{\downarrow}$, currents, at $\hbar \omega=0.2$ in the absence, $u_{1,2}=0.0$, and presence, $u_{1,2}=0.1$, of electron-electron interactions for two complex Floquet driving models. (a) and (b) use model 1 in which Floquet driving for spin-up and spin-down are the same as given by Eq.~(\ref{eq:67R}). (c) and (d) use model 2, in which $[h]_{\downarrow}(t)=[h]_{\uparrow}(t)^*$. Here, $A=0.1/\sqrt{2}$, $\Gamma =0.005$, and the rest of the parameters are the same as Fig.~\ref{fig:2}. } 
\end{figure}
Interestingly, we can see that the complex Floquet driving in model 2 results in a voltage-dependent polarized current, see Figs.~\ref{fig:7R}(c) and \ref{fig:7R}(d). This shows that the polarized light can introduce spin current without any spin-orbit couplings.
In addition, model 1 does not make a substantial modification to the current characteristics as compared to the non-Floquet case, see the trivial small dents in Figs.~\ref{fig:7R}(a) and \ref{fig:7R}(b). 
In Fig.~\ref{fig:7R}(b), we can see that the number of major plateaus doubled such that the onset of the second plateaus (charge quantization) shifts by the values $u_{1,2}$ in respect to the plateaus in Fig.~\ref{fig:7R}(a). The height ratio between the first and the second plateau is almost 2:1 consistent with non-Floquet case \cite{zimbovskaya2008electron}.
One can also see that the interplay between the electron-electron interaction and the Floquet driving in model 2 affects the spin-up and spin-down currents in different ways. 

\section{Conclusion}
In the present work, we have derived four full Floquet-based formulations for quantum transport through a time-periodic dot. In particular, we have shown the two Floquet Green's function formalisms rely on the expansion of Green's function, the first Floquet quantum master formalism, HS-FQME, relies on the Floquet time evolution operator, and the second Floquet quantum master formalism, FS-FQME relies on expanding the Hilbert space to the Floquet space. 
In addition, we have connected the equation of motion for each formulation to the time average observables. 
Through a quantum dot model Hamiltonian, we quantitatively demonstrated that all Floquet quantum transport methods 
{\color{black} 
have a good agreement with each other}
for non-interacting Hamiltonian in the weak coupling regime. 
The frequency and amplitude of time-periodic driving can significantly manipulate the time-average of observables.
We have found that an intense light can shift the one-set and the number of quantized plateaus in the voltage-current curve provided choosing an appropriate light frequency. 
{\color{black} 
For a spin system with the electron-electron term, Floquet driving results in tiny quantized plateaus. However, if the electron-electron interaction strength exceeds the Floquet driving energy then the first charging quantized plateaus will survive.}
{\color{black} 
In addition, we show that a certain complex Floquet driving can result in a polarized time average current.}
As a drawback (even with parallel computation), we found Floquet quantum master equation frameworks and in particular Floquet space Floquet QME to be computationally expensive.
We also found that Floquet QME methods need much smaller time steps compared to the conventional QME. As usual with QME, when the lead-dot coupling or temperature decreases, the number of time steps has to be increased. 
While the QME is a practical method for interacting systems in the weak regime, it is computationally challenging to employ the full Floquet quantum master equation formalisms for large many-body systems. We believe the power of Floquet NEGF and Floquet QME methods goes well beyond the considered example and one needs to combine Floquet QMEs with newly developed machine learning methods to fully benefit from them.  
\begin{acknowledgments}
This material is based upon work supported by the National Natural Science Foundation of China and (NSFC No. 22273075). V.M. acknowledges funding from the Summer Academy Program for International Young Scientists (Grant No. GZWZ[2022]019). W.D. acknowledges the startup funding from Westlake University. V.M. also thanks Rui-Hao Bi and Yong-tao Ma for their technical help.
\end{acknowledgments}
\section*{Conflict of interest}
There are no conflicts to declare.
\section*{Data availability statement}
The data that support the findings of this study are available
upon reasonable request from the authors.
\bibliography{BibFNEGFQME_R2.bib}
\end{document}